# Localized bioconvection of *Euglena* caused by phototaxis in the lateral direction

N. J. Suematsu[1,2](*), A. Awazu[1], S. Izumi[1], S. Nakata[1], H. Nishimori[1,2]

1 *Graduate School of Science, Hiroshima University, 1-3-1 Kagamiyama, Higashi-Hiroshima*

*739-8526, Japan*

2 *Meiji Institute for Advanced Study of Mathematical Sciences (MIMS), 1-1-1 Higashimita, Tamaku, Kawasaki 214-8571, Japan*

*Euglena*, a swimming micro-organism, exhibited a characteristic bioconvection that was localized at the center of a sealed chamber under bright illumination to induce negative phototaxis. This localized pattern consisted of high-density spots, in which convection was found. These observations were reproduced by a mathematical model that was based on the phototaxis of individual cells in both the vertical and lateral directions. Our results indicate that this convection is maintained by upward swimming, as with general bioconvection, and the localization originates from lateral phototaxis.

Suspensions of various swimming micro-organisms often form a macroscopic ordered pattern called "bioconvection". This spatio-temporal pattern is formed under an external stimulus such as a gradient in the oxygen concentration, light illumination, or gravity, which induces a directional swimming of micro-organisms, *i.e.*, taxis. There have been some reports on types of taxis that induce bioconvection, *e.g.*, chemotaxis [1-3], phototaxis [4] and gravitaxis [5,6]. In a typical explanation of bioconvection, an ordered pattern originates from growth of fluctuations in the number density at an accumulation layer near a surface, which is formed by the upward swimming of micro-organisms [7-10]. In this scenario, taxis in the lateral direction is neglected.

*Euglena gracilis*, a unicellular flagellate, exhibits both positive and negative phototaxis depending on the light intensity [11-13]. When the light intensity is stronger than a critical value (0.2 kW·m$^{-2}$), the cells (*Euglena*) tend to swim away from the light source, otherwise the cells swim toward the light source [12]. Furthermore, the cells are sensitive to a gradient of light intensity: the cells assemble in the weakly-illuminated region, although the bright region is not a light source [13].

Robbins reported that a spot pattern with three-fold symmetry is formed upon the bioconvection of *Euglena* throughout an entire chamber with strong light illumination from the bottom of the suspension [4]. In this case, the cells swim away from the bottom light, *i.e.*, upward swimming, because of negative phototaxis in the vertical direction. However, phototaxis in the lateral direction is not remarkable in this pattern, probably because of the high number density and the effect of a surface tension gradient, which is called the "Marangoni effect" [14]. In general, the Marangoni effect is most remarkable for lateral behavior near the surface in the formation of convection [15].

In this paper, we report a novel bioconvection pattern of *Euglena*, *i.e.*, a localized pattern. In our system, a suspension was prepared in a sealed container, where the surface effect was negligible, and illuminated from below with a strong light to induce negative phototaxis. The mechanism of the formation of localized bioconvection was investigated both experimentally and by numerical calculation.

*Euglena gracilis* Pringsheim var. Z (wild-type, strain Z) was kindly provided by Prof. N. Suzaki (Kobe University, Japan). *Euglena* was pre-cultivated statically in 100 mL of Koren and Hutner medium [16] adjusted to pH 6.9. All cultures were inoculated with $5 \times 10^4$ cells·mL$^{-1}$. The cells were grown at 27°C with continuous illumination unless otherwise specified. After saturation, the cells were inoculated into Hyponex aqueous medium (1 g of Hyponex (N-P-K: 6.5-6-19) dissolved in 1 L of distilled water) and grown at room temperature (26±1°C). The number density of the cells $\rho$ was reached $2.5 \times 10^5$ cells·mL$^{-1}$ within a week after inoculation.

A dense suspension was prepared using centrifugation ($\rho \sim 15 \times 10^5$ cells·mL$^{-1}$), and diluted to the desired number density ($1\sim15\times10^5$ cells·mL$^{-1}$). The diluted suspensions were placed an original sealed container. The container consisted of two slide glasses and a silicon pad (thickness: 0.5, 1, 2, and 5 mm) as a spacer. The silicon pad had a hole into which the suspension was poured. Thus, the suspension depth $d$ was equal to the thickness of the silicon pad. A sample was prepared by sealing off the suspension with the two glass plates.

The sealed suspension was illuminated from both the top and bottom. The light sources were a fluorescent light and a flat panel light (Light Viewer 7000 PRO) with intensities of 600 and 3000 Lx, respectively. The macroscopic bioconvection pattern was photographed every 30 sec using a digital camera (EOS Kiss Digital N, Canon). The digital images were analyzed using "ImageJ 1.41" (National Institutes of Health, USA).

A suspension of *Euglena* exhibited dynamic spatio-temporal patterns under illumination from below with a strong light (3000 Lx) and from the top with a relatively-weak light (600 Lx). The patterns consisted of green-coloured spots or lines containing a high number density of *Euglena*. The time-evolution of a typical bioconvection in our experiments is shown in fig. 1a, where $d$ was 2 mm. In the initial condition, the number density was homogeneous throughout the entire system (0 min). After 5~8 minutes, the cells (*Euglena*) assembled and formed high-density spots at random positions. We hereafter refer to these spots as "clusters". These clusters moved away from the silicon wall and gathered at the center of the circular container to form an ensemble of clusters (12~18 min). Outside of an ensemble, the number density was nearly zero. This localized pattern was dynamic, which means that the clusters repeatedly underwent fusion and division. Furthermore, an ensemble of clusters migrated in a group (20~180 min). This process was clearly observed in a space-time diagram of vertical slices at the center of the container (fig. 1b). The features of the pattern were evaluated in terms of the number of clusters $n$ and the distance between the centers of neighboring clusters $\lambda$ over time (fig. 1c). The number of clusters increased with time and was saturated in about 10 minutes. On the other hand, the distance $\lambda$ initially had a large value and decreased with time. This distance eventually reached a constant value of about 2 mm, which was close to the depth of the suspension $d$.

   The stability of the localized pattern was examined by changing the conditions of light illumination. The localized pattern shown in fig. 1 was maintained for 5 hours. If the bottom light was turn off, the localized pattern disappeared within 10 sec, and a large flat aggregate was left. Without the top light, the localized pattern was still formed, but the duration of the pattern decreased to about 3 hours. On the other hand, when the light intensity was reversed, *i.e.*, the strong light on top and the relatively-weak light on the bottom, the localized pattern was not generated, and a large aggregate was only observed at the center of the container.

   To clarify the hydrodynamic behavior within the localized pattern, the motion of individual cells was observed by magnifying around the cluster. Near the upper glass plate, the cells swam toward the center of the cluster. In contrast, at the bottom of the container, the cells spread from the center of the cluster. Therefore, there might be downward flow of the cells at the center of cluster.

To characterize the features of the localized pattern, $n$ and $\lambda$ were measured while varying the suspension depth $d$ and the number density of the suspension $\rho$. Since $n$ and $\lambda$ changed with time and reached constant values within several tens of minutes, these values were measured in 60 minutes after starting the illumination. The average value of $\lambda$ increased with an increase in $d$ (fig. 2a,b). Here, $\lambda=0$ indicates a single cluster or a flat aggregate. When $d$ was 5 mm, a line or roll pattern was observed. In contrast to $d$, $\rho$ had less of an effect on $\lambda$, but was positively correlated with $n$ (fig. 2c). Namely, the area occupied by the localized pattern became large with an increase in $\rho$.

We compared the formation process of a localized pattern with that of a general bioconvection pattern. Bioconvection is generally considered to be a hydrodynamic behavior similar to Rayleigh-Bénard convection [7-9]. When an external stimulus induces the upward swimming of micro-organisms, a top-heavy condition is generated because of the accumulation of heavy micro-organisms near the surface of the suspension. If the gradient of weight density is high enough to become the fluid unstable, sedimentation of the cells occurs and convection is induced all over the chamber [10]. In contrast to general bioconvection, *Euglena* firstly generates clusters at random positions. This early process appears to be more similar to nuclear formation and growth rather than the growth of fluctuation. In addition, after this early process, a localized pattern is formed as a result of the gathering the clusters. The localized pattern is distinctly different from general bioconvection pattern that appears in whole of a container. Although Yamamoto *et al.* reported a localized bioconvection of *chlamydomonas reinhardtii* similar to our result, they only mentioned that the localization was caused by inhomogeneity of light intensity [17]. We consider that the localization of the bioconvection originate from a property of *Euglena*. The formation processes of localized bioconvection reveal that the cells tend to swim toward a high-density region in the lateral direction. This tendency is supported by the aggregation of cells under strong light illumination from above. Thus, it is supposed that the phototaxis of *Euglena* in the lateral direction plays an important role in pattern formation in our system. Here, we propose two types of scenarios of the phototaxis in the lateral direction. One is a self-shading effect [18]. Since a high-density region shades a light from below, the cells tend to assemble to avoid bright illumination. The other is caused by green-coloured light scattered from the cells themselves. Namely, a high-density region plays a roll of light source to induce the positive phototaxis in the lateral direction.

Phototaxis in the vertical direction is needed to maintain the bioconvection of *Euglena*.

Without a high-intensity light from below, the clusters and the localized pattern either disappear or are not formed. These results indicate that the formation of clusters requires upward swimming. In addition, a cluster consists of convective flow, as shown under magnification. Therefore, the behavior of individual cells around the cluster is similar to that in general bioconvection. Namely, upward swimming induces a top-heavy condition, and aggregates of heavy cells then drop down due to their own weight.

If we consider that the distance $\lambda$ corresponds to the wavelength of the convection pattern, the positive correlation between $\lambda$ and $d$ agrees with that in thermal convection. It has been shown that a characteristic wavelength in a thermal convection pattern increases with the Rayleigh number, which increases with the depth of the medium [15]. Childress *et al*. [8] and Levandowsky *et al*. [9] introduced a dimensionless number to bioconvection, which is positively correlated with both the suspension depth and the number density, similar to the Rayleigh number. However, the number density of a suspension $\rho$ does not influence $\lambda$ in our experiments (fig. 2c). This is because the partial number density changes with the lateral assembly of cells and is independent of $\rho$. To compensate for the independence of $\lambda$, $n$ increases with an increase in $\rho$.

The characteristic pattern of bioconvection of *Euglena* is considered to originate from phototaxis in the lateral direction, although convection is still caused by upward swimming. To support this notion, we constructed a mathematical model based on the assumption of phototaxis in both the vertical and lateral directions. Although the mathematical model for bioconvection of phototactic micro-organisms, especially the effect of the self-shading, has already been reported by S. Ghorai and N. A. Hill [18], the generation of the localized pattern has not yet been succeeded. In contrast, we consider the effect of phototaxis in the lateral direction.

We consider a quasi-one-dimensional system with length $L$ which consists of two layers, *i.e.*, upper and lower. We made three assumptions based on the phototactic behavior of individual cells: (i) The rate of the upward transition is greater than that of the downward transition, since the cells swim away from the strong illumination at the bottom. (ii) Swimming in the lateral direction tends to be toward a high-density region in the upper layer, due to positive phototaxis in the lateral direction. Thus, the direction of swimming tends to turn toward the source of green light that is scattered by the cell itself. Since the intensity of the scattered light might be weaker than the critical value, the cell exhibits positive phototaxis. (iii) The rate of the downward transition increases with an increase in the total population $P$, which is the sum of the populations in both the upper and lower layers, since the weight of a dense aggregate increases with $\rho$.

The time-evolution of the populations in the upper and lower layers can be described by the

following equations:

$$\frac{\partial P^u}{\partial t} = -\frac{a}{\Delta}\left[\frac{P}{\Delta}\right]^\gamma P^u + \frac{b}{\Delta}P^l + D_u\frac{\partial^2 P^u}{\partial x^2} - \frac{\partial}{\partial x}\left[(C_+ - C_-)P^u\right], \quad (1)$$

$$\frac{\partial P^l}{\partial t} = \frac{a}{\Delta}\left[\frac{P}{\Delta}\right]^\gamma P^u - \frac{b}{\Delta}P^l + D_l\frac{\partial^2 P^l}{\partial x^2}, \quad (2)$$

$$C_+ = c\int_x^\infty e^{-\frac{|x-y|}{\Lambda}}\frac{P^u}{\Delta}dy, \quad C_- = c\int_{-\infty}^x e^{-\frac{|x-y|}{\Lambda}}\frac{P^u}{\Delta}dy, \quad (3)$$

where $P^u$ and $P^l$ are the cell populations in the upper and lower layers, $P$ is the total population ($P=P^u+P^l$), $a$ and $b$ are constants that correspond to average velocities in the downward and upward directions, $\Delta$ is the depth of the suspension, $D_u$ and $D_l$ are the diffusion coefficients in the upper and lower layers, $c$ is a coefficient which indicates the relationship between the number density of the cells and the light intensity, and $\Lambda$ is the decay length of light propagation. In both eqs. (1) and (2), the first and second terms reflect the transition of cells toward the lower and upper layers, respectively. The value of $a$ is assumed to be much smaller than that of $b$, due to assumption (i). In addition, the rate of the downward transition (first term) depends on the partial number density, i.e., $P/\Delta$, due to assumption (iii). The final term in eq. (1) reflects positive phototaxis in the lateral direction, as explained in assumption (ii). Equation (3) defines $C_+$ and $C_-$, which are related to phototaxis toward the right (+) and left (−) directions. Here, we consider that the intensity of light scattered from the cells is proportional in the number density and decays with the distance |x − y|, which is based on Lambert-Beer law [19]. The average number density in space is given by $Q = \frac{1}{L\Delta}\int_0^L P(x)dx$.

We performed a simulation of this model for some values of $Q$ and $\Delta$ with the fixed values $a = 1$, $b = 1000$, $c = 300$, $L = 1$, $\gamma = 2$, $\Lambda = 1/16$, $D_u = 2$, $D_l = 1/25$. The boundary condition of $P^u$ and $P^l$ is the Neumann boundary condition and that of $C_+$ and $C_-$ is calculated with $P^u = 0$ at both $x < 0$ and $x > L$, since there are no cells outside the system. In the initial condition, the population is uniform the throughout the entire system.

The localized pattern is reproduced by the numerical calculation. The process of pattern formation is shown in the space-time diagram of $P^l$ (fig. 3a,b). The diagrams indicate that cells assemble in the initial state, and then high-population ($P^l$) regions, which we call "clusters", are generated, and this is followed by the gathering of clusters. As a result, a periodic pattern in space is formed by the population profile. Cluster formation and gathering is observed in fig. 3a,b, where the depth $\Delta$ in fig. 3a is less than that in fig. 3b. If we focus on a single cluster, convection-like behavior is observed. Figure 3c shows the upward transition rate ($F^v$, Top), horizontal flow in the upper and lower layers ($F^u$ and $F^l$, Middle), and populations in the upper and lower layers (Bottom). The transition rate and horizontal flow are calculated with the following equations:

$$F^v = -\left[\frac{P}{\Lambda}\right]^\gamma \frac{a}{\Lambda} P^u + \frac{b}{\Lambda} P^l, \tag{4}$$

$$F^u = (C_+ - C_-)P^u - D_u \frac{\partial P^u}{\partial x}, \tag{5}$$

$$F^l = -D_l \frac{\partial P^l}{\partial x}. \tag{6}$$

The profile of $F^u$ indicates flow toward the center of the cluster in the upper layer, and a negative value of $F^v$ at the center of the cluster indicates a downward transition. The profile of $F^l$ then shows flow away from the center of the cluster. Furthermore, the positive value of $F^v$ around the cluster indicates an upward transition. Therefore, there is convective flow around the cluster. The number of clusters and the cluster distance depend on both $\Delta$ and $Q$. With an increase in $\Delta$, the number of clusters within the same region decreases (compare the top and bottom profiles with the same value of $Q$ in fig. 3d). This result indicates that the cluster distance increases with an increase in $\Delta$. On the other hand, with an increase in $Q$, the cluster distance remains constant and the number of clusters increases. These relationships are independent of the initial conditions.

The results indicate that the numerical calculation qualitatively reproduces the experimental results: *i.e.*, the process of pattern formation (fig. 1), convection around a cluster, and the dependency of $n$ and $\lambda$ on $d$ and $\rho$ (fig. 2a,b). The mathematical model is based on assumptions regarding the phototaxis of individual cells in both the vertical and lateral directions. Therefore, the numerical calculation supports our supposition that the localization of the bioconvection of *Euglena* is caused by phototaxis in the lateral direction.

In conclusion, a localized bioconvection pattern was generated in a suspension of *Euglena* in a sealed container under light illumination from below. The localized pattern consisted of clusters where convective flow is found. The cluster distance $\lambda$ increased with the suspension depth $d$, and the number of clusters $n$ increased with the number density $\rho$. The results of the experiment and the numerical calculation indicate that convection is maintained by upward swimming, and localization of the pattern is due to phototaxis in the lateral direction.

**Acknowledgment**

We thank Prof. S. Kitsunezaki (Nara Women's University, Japan) for his helpful discussion regarding bioconvection, Prof. N. Suzaki (Kobe University, Japan) for providing technical advice on how to grow *Euglena*, and Mr. Y. Iwata (Hiroshima University, Japan) for his assist to culture the Euglena. This work was supported in part by Meiji university Global COE Program "Formation and Development of Mathematical Sciences Based on Modeling and Analysis" from the MEXT of Japan.

* suematsu@hiroshima-u.ac.jp

Figures

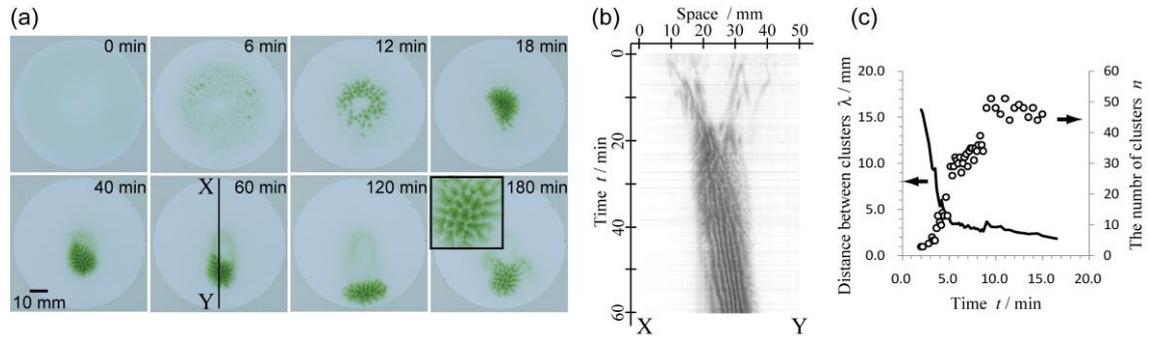

Fig. 1: (a) Snapshots during the process of pattern formation by *Euglena*. The diameter of the circular container was 50 mm, and the suspension depth was 2 mm. (b) A space-time diagram of vertical slices at the center of the container described by the X-Y line in the snapshot at 60 min. The diagram shows the distribution of the number density with a gray scale where the black region indicates a high number density. (c) Time-evolution of the distance between clusters $\lambda$ and the number of clusters $n$.

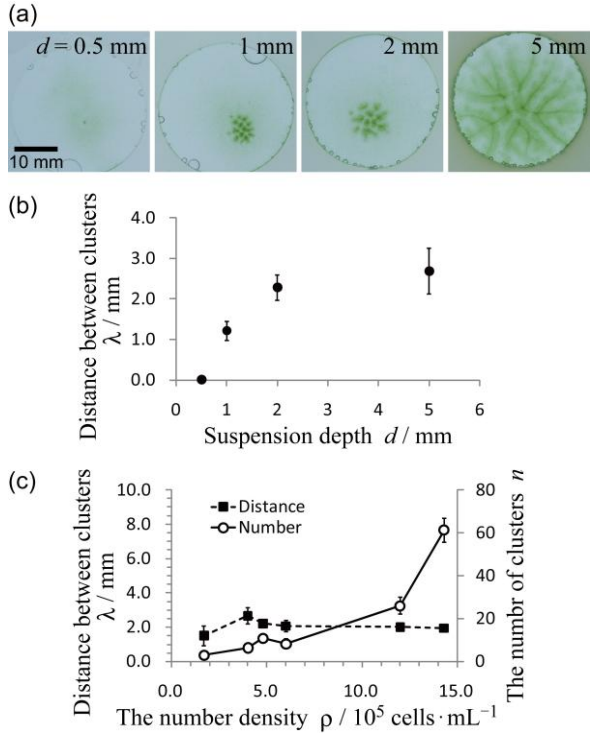

Fig. 2: (a) Snapshots of bioconvection depending on the suspension depth $d$. (b) Distance between the centers of neighboring clusters $\lambda$ depending on $d$. (c) The number of clusters $n$ (○) and $\lambda$ (■) depending on the number density $\rho$ with 2 mm in $d$. All values of $n$ and $\lambda$ were estimated after 60 minutes of starting the light illumination.

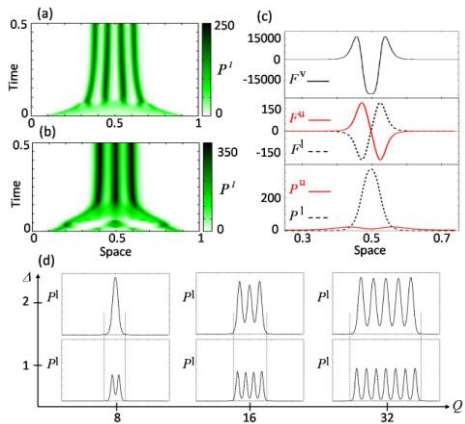

Fig. 3: Results of the numerical calculation. (a)(b) Space-time diagram of the population in the lower layer $P^l$ with $Q=16$ and (a) $\varDelta=1$, (b) $\varDelta=2$. (c) Profiles of (top) transition rate $F^v$, (middle) horizontal flow in the layers of upper $F^u$ and lower $F^l$, and (bottom) $P^u$ and $P^l$ around a single cluster in which $Q=8$ and $\varDelta=2$. (d) Phase diagram of the localized pattern of $P^l$ depending on $\varDelta$ and $Q$.